**Title**

Characterizing cycle structure in complex networks

**Authors**

Tianlong Fan[1,2,3], Linyuan Lü[1,3,4,7,*], Dinghua Shi[5,*], Tao Zhou[6,*]

**Summary**

Cycle is the simplest structure that brings redundant paths in network connectivity and feedback effects in network dynamics. Focusing on cycle structure, this paper defines a new matrix, named cycle number matrix, to represent cycle information of a network, and an index, named cycle ratio, to quantify the node importance. Experiments on real networks suggest that cycle ratio contains rich information in addition to well-known benchmark indices, for example, the node rankings by cycle ratio are largely different from rankings by degree, H-index, coreness, betweenness and articulation ranking, while the rankings by degree, H-index, coreness are very similar to each other. Extensive experiments on identifying vital nodes that maintain network connectivity, facilitate network synchronization and maximize the early reach of spreading show that cycle ratio is competitive to betweenness and overall better than other benchmarks. We believe the in-depth analyses on cycle structure may yield novel insights, metrics, models and algorithms for network science.

**Introduction**

The last two decades have witnessed extensive development in network science (Newman, 2018), with research focuses being shifted from discovering macroscopic properties (Barabási and Albert, 1999; Newman, 2002; Watts and Strogatz, 1998) to uncovering the functional roles played by microscopic structures, or even individual nodes and links (Alon, 2007; Lü et al., 2016a; Lü and Zhou, 2011). Scientists have pieced an increasingly clear picture about the functions of specific structures in disparate dynamical processes, such as the roles of different motifs in biological and communication networks (Alon, 2007), how information and behaviors propagate along a contacting chain (Christakis and Fowler, 2007), and how a local star structure self-sustains an epidemic spreading process (Castellano and Pastor-Satorras, 2010).

[1]Institute of Fundamental and Frontier Sciences, University of Electronic Science and Technology of China, Chengdu 611731, China.
[2]Department of Physics, University of Fribourg, Fribourg 1700, Switzerland
[3]Alibaba Research Center for Complexity Sciences, Alibaba Business College, Hangzhou Normal University, Hangzhou 311121, China.
[4]Beijing Computational Science Research Center, Beijing 100193, China.
[5]Department of Mathematics, Shanghai University, Shanghai 200444, China.
[6]CompleX Lab, University of Electronic Science and Technology of China, Chengdu 611731, China.
[7]Lead Contact.
*Correspondences: L.L. (linyuan.lv@gmail.com), D.S. (shidh2012@sina.com) and T.Z. (zhutou@ustc.edu).

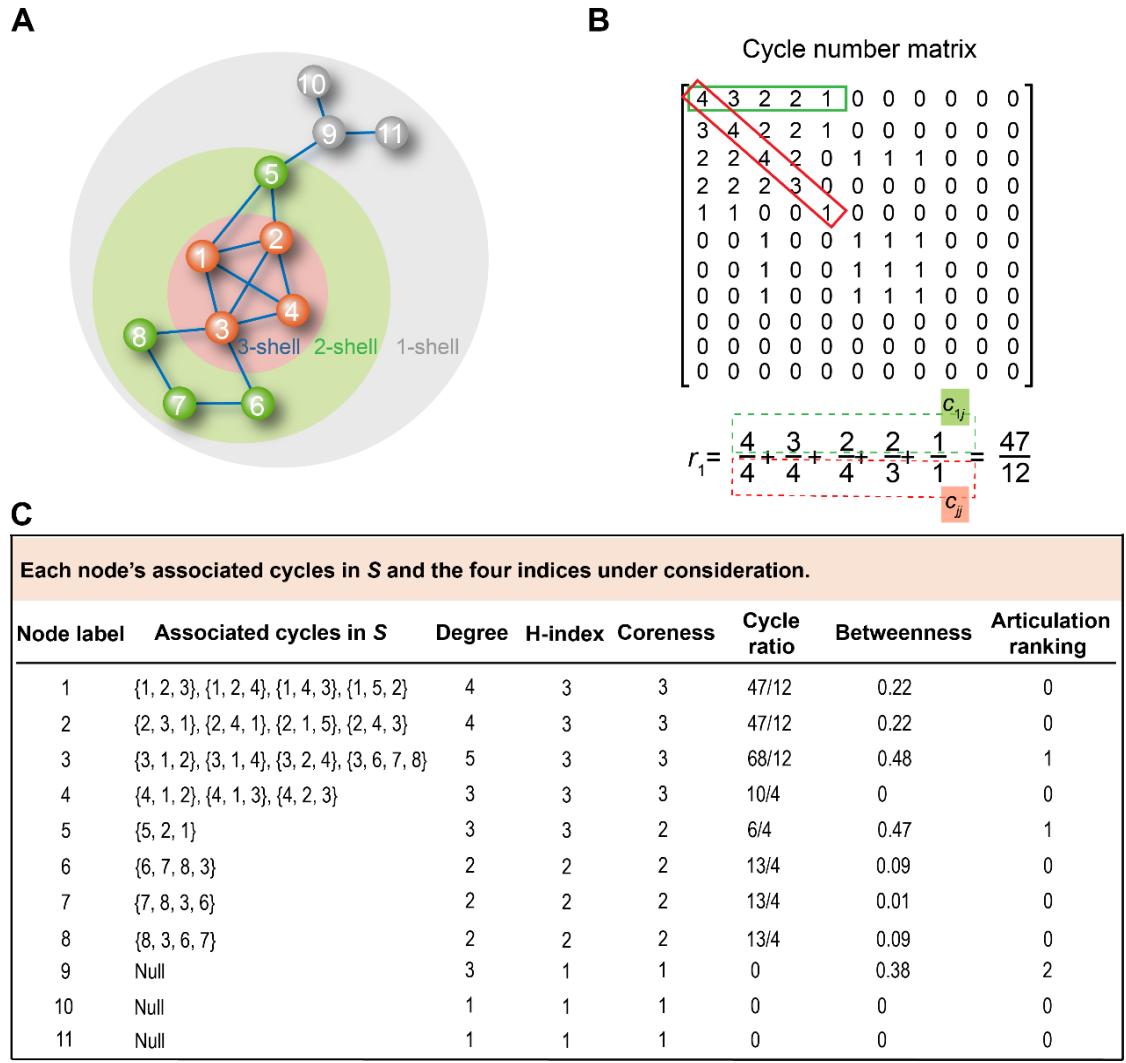

Fig. 1. Cycle ratios of nodes in an example network. (A) An example network which has three shells according to the *k*-core decomposition (Kitsak et al., 2010). (B) The corresponding cycle number matrix and how to calculate the cycle ratio of node 1. (C) Every node's associated cycles in *S*, degree, H-index, coreness, cycle ratio, betweenness and articulation ranking. For this example network, the set of shortest cycles is $S = \{\{1,2,3\}, \{1,2,4\}, \{1,2,5\}, \{1,3,4\}, \{2,3,4\}, \{3,6,7,8\}\}$.

Besides extensively studied chain and star structures, cycle is another ubiquitously observed structure (Kim and Kim, 2005), which plays significant roles in both structural organization and functional implementation. A cycle, also called loop in literature, can be simply defined as a closed path with the same starting and ending node. Recent studies have uncovered the topological properties of cycles, including the distribution of cycles of different sizes in real and artificial networks (Bianconi et al., 2008, 2005; Bianconi and Capocci, 2003; Bonneau et al., 2017; Rozenfeld et al., 2005), the effect of degree correlations on the loops of scale-free networks (Bianconi and Marsili, 2006), as well as the significant roles of the cycles in network functions related to storage (Lizier et al., 2012), synchronizability (Shi et al., 2013) and controllability (Ruths and Ruths, 2014). In addition, the organization of cycles can be utilized to characterize individual nodes and links. For example, a measure called clustering coefficient (Watts and Strogatz, 1998) is based on counting the number of associated triangles (triangle is the cycle with smallest size), which was recently extended to account for the associated cycles with larger sizes (Caldarelli et al., 2004; Fronczak et al., 2002; Kim and Kim, 2005), the effect of the addition of a none-observed link on the local organization of cycles can be used to estimate the likelihood of the existence of this link (Pan et al., 2016), and the probability a self-avoid random walker returns to the target node through a cycle (cycles with different lengths are assigned to different weights) can be used to quantify the importance of the target node (Van Kerrebroeck and Marinari, 2008).

Considering a simple network where direction and weight of a link are ignored and self-loops are not allowed, then a cycle is the simplest structure providing redundant paths to all involved node pairs. That is to say, if two nodes belong to a cycle, there are at least two independent paths connecting them. Such redundancy also brings complicated feedbacks in interacting dynamics. Therefore, the in-depth understanding of cycle structure may provide insights and methods on how to maintain the network connectivity under attacks (Albert et al., 2000) and how to regulate interacting dynamics towards predesigned states (Arenas et al., 2008).

In this paper, according to the cycle-based statistics, we propose a novel matrix (named cycle number matrix) to represent cycle information of network, and a new index (named cycle ratio) to quantify the importance of individual nodes. This index is essentially different from well-known indices and methods (Lü et al., 2016a), producing a much different ranking of nodes comparing with degree (Barabási and Albert, 1999), H-index (Lü et al., 2016b), coreness (Kitsak et al., 2010) and betweenness (Brandes, 2001). Extensive experiments on real networks in identifying the most vulnerable nodes under intentional attacks (Callaway et al., 2000; Cohen et al., 2001), the most efficient nodes in pinning control (Li et al., 2004; Wang and Chen, 2001; Qiu et al., 2021) and the most influential nodes in the early stage of epidemic spreading (Pastor-Satorras et al., 2015; Zhou et al., 2019) show that cycle ratio is competitive to betweenness and overall better than other benchmarks including degree, H-index, coreness and articulation ranking (Tishby et al., 2018).

**Results**

Considering a simple network $G(V, E)$, where $V$ and $E$ are the sets of nodes and links, respectively. The size of a cycle equals the number of links it contains. The cycles containing node $i$ with the shortest size are defined as node $i$'s associated shortest cycles (also called $i$'s shortest cycles for simplicity) and the corresponding size is called node $i$'s girth (Shi et al., 2013). Denote by $S_i$ the set of the shortest cycles associated with node $i$, and $S = \cup_{i \in V} S_i$ the set of all shortest cycles of $G$, we define the so-called cycle number matrix $C=[c_{ij}]_{N \times N}$ to characterize the cycle structure of $G$, where $N=|V|$ is the number of nodes in $G$, and $c_{ij}$ is the number of cycles in $S$ that pass through both nodes $i$ and $j$ if $i \neq j$. If $i = j$, $c_{ii}$ is the number of cycles in $S$ that contain node $i$. Obviously, $C$ is a symmetric matrix. On the basis of the cycle number matrix, we propose an index, named cycle ratio, to measure a node's importance as

$$r_i = \begin{cases} 0, & c_{ii} = 0 \\ \sum_{j, c_{ij} > 0} \frac{c_{ij}}{c_{jj}}, & c_{ii} > 0 \end{cases}. \quad (1)$$

According to the above definition, if a node $i$ doesn't belong to any cycle in $S$, its cycle ratio is reasonably set to be zero. When $c_{ii} > 0$, all items in the summation are well defined since $c_{jj} > 0$ if $c_{ij} > 0$. The ratio estimates the importance of node $i$ subject to its participation to other nodes' shortest cycles in $S$. Note that, in our definition, only shortest cycles associated with each node are considered since cycles with larger sizes are usually less relevant to the network functions (we have also tested on longer cycles, see details in Discussion) and to account for all cycles is infeasible for most networks due to the tremendous computational complexity (Pan et al., 2016) (Fig. S1 in Supplementary Information shows the number of cycles with different lengths, indicating an exponential growth). Fig. 1A presents an example network, and Fig. 1B shows the corresponding cycle number matrix. The process to calculate the cycle ratio of an example node (i.e., node 1) is also shown in Fig. 1B. In Equation 1, each term represents the degree to which node $i$ ($i$=1 for this example) participates in $j$'s associated shortest cycles ($j = 1, 2, 3, 4$ and $5$ for this example) in which denominator is the number of shortest cycles of node $j$, and the numerator is the number of cycles associated with both node $i$ and node $j$. For example, the second term in the example equation in Fig. 1B, 3/4, means that three of the four shortest cycles of node 2 ({2, 3, 1}, {2, 4, 1}, {2, 1, 5}, {2, 4, 3}) contain node 1. In a word, $r_1$ represents the degree to which node 1 participates in associated shortest cycles of other nodes. The cycle ratios of all nodes are presented in Fig. 1C. Five well-known node centralities, degree (Barabási and Albert, 1999), H-index (Lü et al., 2016b), coreness (Kitsak et al., 2010), betweenness (Brandes, 2001) and articulation ranking (Tishby et al., 2018) (see precise definitions of these indices in Methods), are used as benchmarks for comparison. Their values for this example network are also presented in Fig. 1C.

**Table 1. The basic topological features of the six networks.** Here $N$ and $M$ are the number of nodes and links, $\langle k \rangle$ and $\langle L \rangle$ are the mean degree and mean shortest distance, and $C$ is the mean clustering coefficient (Watts and Strogatz, 1998).

| Network | $N$ | $M$ | $\langle k \rangle$ | $\langle L \rangle$ | $C$ |
|---|---|---|---|---|---|
| **C. elegans** | 297 | 2148 | 14.46 | 2.46 | 0.29 |
| **Email** | 1133 | 5451 | 9.62 | 3.61 | 0.22 |
| **Jazz** | 198 | 2742 | 27.70 | 2.24 | 0.62 |
| **NS** | 379 | 914 | 4.82 | 6.04 | 0.74 |
| **USAir** | 332 | 2126 | 12.81 | 2.74 | 0.63 |
| **Yeast** | 2375 | 11693 | 9.85 | 5.10 | 0.31 |

We test the performance of cycle ratio in identifying vital nodes subject to three well-studied dynamical processes, node percolation (Callaway et al., 2000; Cohen et al., 2001), synchronization (Arenas et al., 2008) and epidemic spreading (Pastor-Satorras et al., 2015). The first one considers nodes' ability to maintain the network connectivity, the second one accounts for nodes' capacity to regulate interacting dynamics towards a certain predesigned state, and the last one concentrates on infected nodes' reach in the early stage of an epidemic outbreak. The experiments are carried out on six real networks from disparate fields, including the neural network of C. elegans (C. elegans) (Rossi and Ahmed, 2015), the email communication network of the University at Rovira i Virgili in Spain (Email) (Guimerà et al., 2003), the collaboration network of jazz musicians (Jazz) (Gleiser and Danon, 2003), the collaboration network of scientists working on network science (NS) (Newman, 2006), the US air transportation network (USAir) (Batagelj and Mrvar, 2006), and the protein-protein interaction network of yeast (Yeast) (Jeong et al., 2001). Their basic topological features are summarized in Table 1.

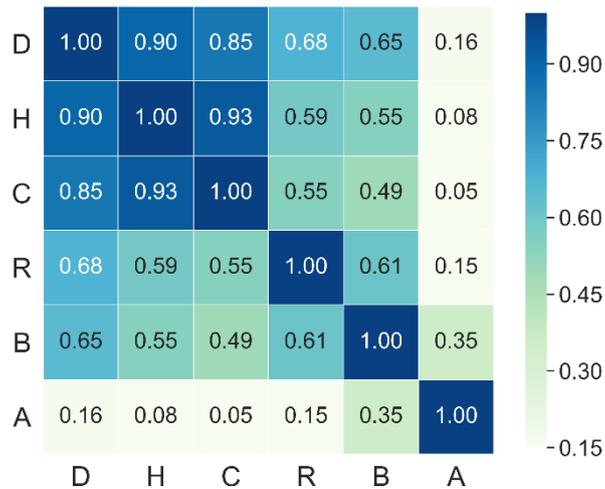

**Fig. 2. The average correlation matrix of the six indices over the six networks.** Here D, H, C, R, B and A represent degree, H-index, coreness, cycle ratio, betweenness and articulation ranking, respectively. Each element is the averaged value of $\tau$ over the six networks, and the value is visualized by the color.

Before penetrating into each index's ability to identify vital nodes, we first see whether cycle ratio contains rich information in addition to the five benchmarks. We apply the Kendall's Tau ($\tau$) (Kendall, 1938; Knight, 1966) to measure the correlation between pairs of indices (see the definition of $\tau$ in Methods). Given two indices *X* and *Y*, if $\tau(X,Y)$ is close to 1, it indicates that *X* and *Y* are highly correlated and less differential to each other. Fig. 2 shows the average correlation matrix between all index pairs for the six networks (the correlation matrix for each network is shown in Supplementary Information Fig. S2), one can clearly observe that the correlations between degree, H-index and coreness are high (the average of $\tau$ between them is 0.89), while the correlations between articulation ranking, cycle ratio, betweenness and other indices are low. That is to say, the resulted node rankings produced by degree, H-index and coreness are very similar to each other. Therefore, although the performance of H-index or coreness in some specific tasks is better than degree (Kitsak et al., 2010; Lü et al., 2016b), the node rankings produced by H-index and coreness contain less information in addition to the one produced by degree, and vice versa. In contrast, as suggested by the lower correlations, the node rankings produced by cycle ratio and betweenness have rich information in addition to these produced by degree, H-index and coreness. This is a very important yet easy-to-be-ignored marker about the potential value of a newly proposed index since the lower correlations between the proposed index and known indices indicate a higher possibility that the proposed index will provide novel insights beyond known indices. Although the correlations between articulation ranking and other indices are the lowest, it is mainly caused by its low distinguishability, and we will see later that in many cases, the performance of articulation ranking is very poor. Supplementary Information Section III shows the distributions of the six indices for the six real networks under consideration. One can observe that the distinguishability of cycle ratio is good while the distinguishabilities of articulation ranking and coreness are poor.

Fig. 3 presents visualized Yeast network corresponding to the resulted rankings by the six indices. It can be seen intuitively that the vital nodes selected by degree, H-index and coreness are densely connected with each other and clustered in a certain region, in consistent to the so-called rich-club phenomenon (Colizza et al., 2006; Zhou and Mondragón, 2004). As a contrast, the vital nodes selected by cycle ratio and betweenness are scattered in the whole network with sparser connections among them. This is a significant advantage of cycle ratio and betweenness if one would like to find out a set of vital nodes, because if the selected vital nodes tend to be clustered to each other, their influential areas will be highly overlapped and thus their collective influences are probably weaker (Lü et al., 2016a; Ji et al., 2017; Kitsak et al., 2010; Zhang et al., 2016). Although the visualizations of cycle ratio, betweenness and articulation ranking in Fig. 3 look similar, the Kendall's Tau between cycle ratio and them are only 0.56 and 0.09, respectively (see Supplementary Information Fig. S2). Therefore, we believe the in-depth analyses of cycle ratio may uncover novel insights that cannot be directly obtained by other benchmark centralities.

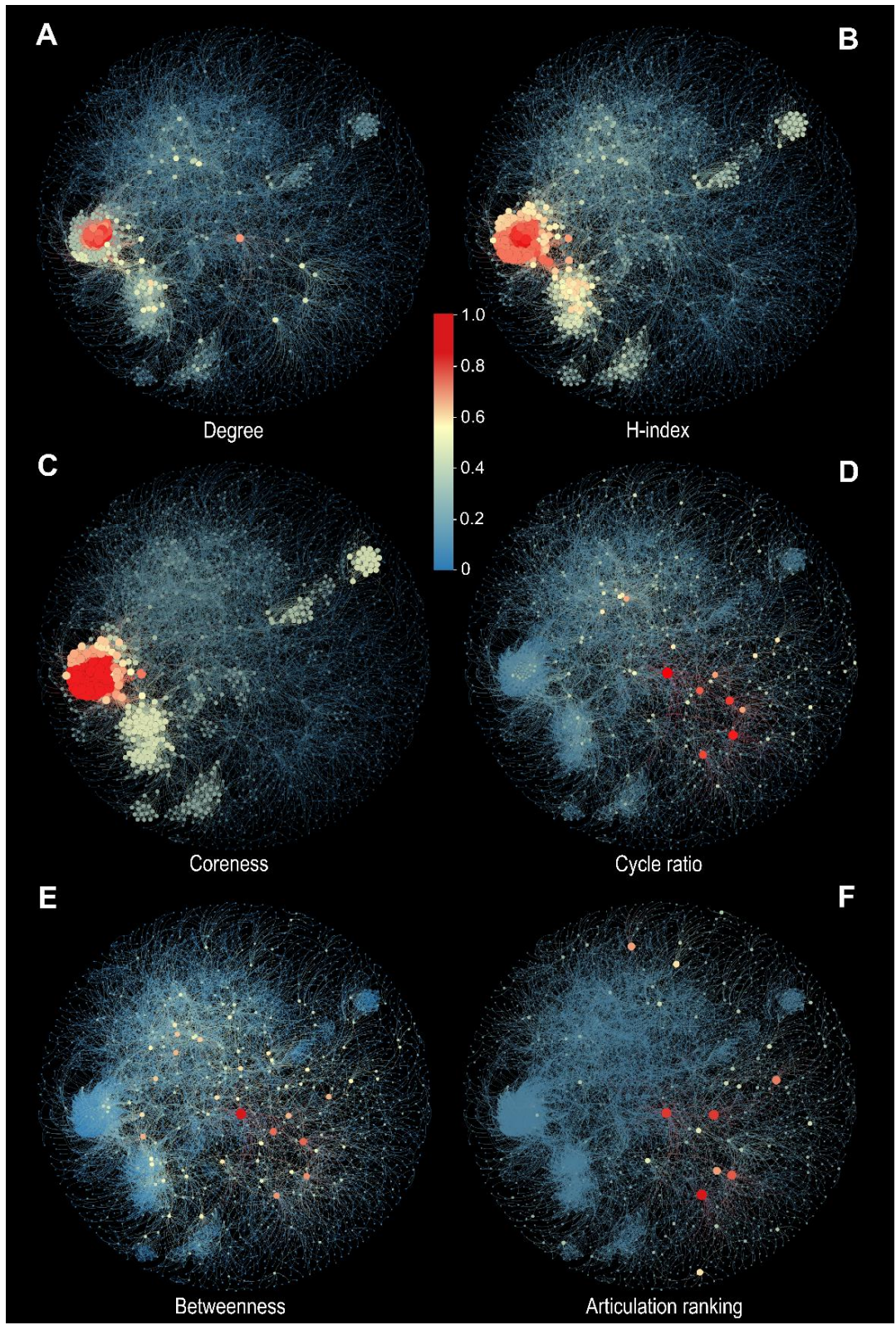

**Fig. 3. Visualization of the rankings of nodes produced by degree, H-index, coreness, cycle ratio, betweenness and articulation ranking.** The Yeast network is taken for example. In each plot, the sizes of nodes are proportional to their values of the corresponding index. The color of a node indicates its relative value normalized by the maximum value. For example, in Plot **A**, a node $i$'s relative value is $k_i/k_{max}$ where $k_i$ is $i$'s degree and $k_{max}$ is the maximum degree of Yeast.

To evaluate the importance of nodes in maintaining the network connectivity, we study the node percolation dynamics (Callaway et al., 2000; Cohen et al., 2001). Given a network, we remove one node at each time step and calculate the size of the largest component of the remaining network until the remaining network is empty. The metric called Robustness (Schneider et al., 2011) is used to measure the performance, defined as

$$R = \frac{1}{N} \sum_{n=1}^{N} g(n), \qquad (2)$$

where the relative size $g(n)$ is the number of nodes in the largest component divided by $N$ after removing $n$ nodes. The normalization factor $1/N$ ensures that the values of $R$ of networks with different sizes can be compared. For each index, the nodes are ranked in the descending order and the ones in the top places are removed preferentially. Obviously, a smaller $R$ means a quicker collapse and thus a better performance. Fig. 4 shows the collapsing processes in the six real networks, resulted from the node removal by cycle ratio and the other five indices. In most cases betweenness leads to faster collapses than others, but its advantage tends to come at later stages. Table 2 exhibits the Robustness $R$, from which one can see that betweenness performs best, cycle ratio is close to the best, and articulation ranking is the worst.

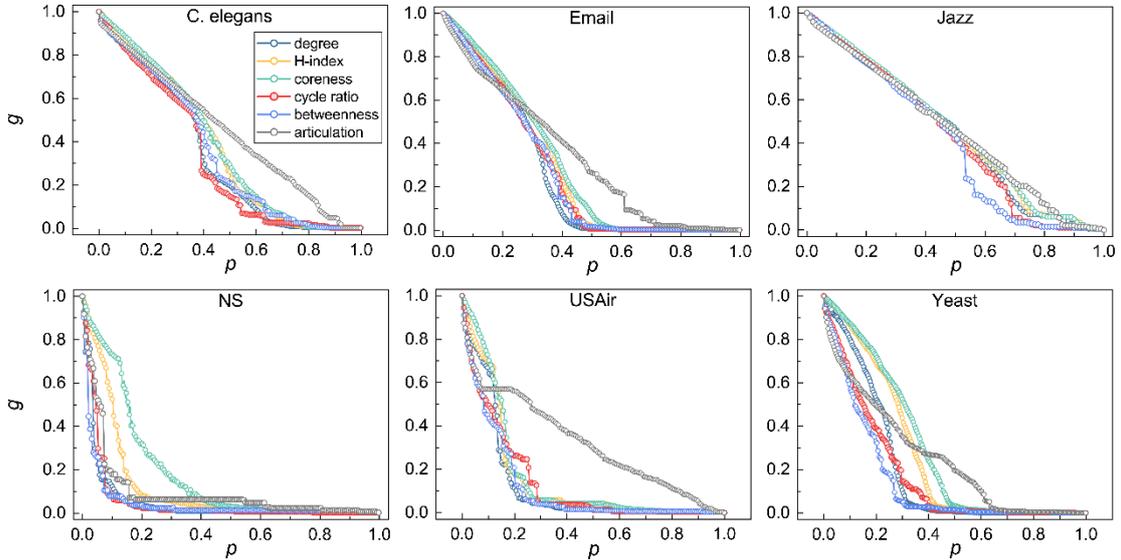

**Fig. 4. The performance of the six indices on node percolation.** The *y*-axis shows the relative size of the largest component after node removal and the *x*-axis denotes the ratio of removed nodes

We next evaluate the importance of nodes by measuring the effect caused by pinning these nodes in a synchronizing process (Li et al., 2004; Wang and Chen, 2001). Considering a general case where a simple connected network $G(V, E)$ is consisted of $N$ linearly and diffusively coupled nodes, with an interacting dynamics as

$$\dot{x}_i = f(x_i) + \sigma \sum_{j=1}^{N} l_{ij} \Gamma(x_j) + U_i(x_i, \ldots, x_N), \tag{3}$$

where the vector $x_i \in \mathbf{R}^n$ is the state of node $i$, the function $f(\cdot)$ describes the self-dynamics of a node, the positive constant $\sigma$ denotes the coupling strength, $U_i$ is the controller applied at node $i$, and the inner coupling matrix $\Gamma: \mathbf{R}^n \to \mathbf{R}^n$ is positive semidefinite. The Laplacian matrix $L = [l_{ij}]_{N \times N}$ of $G$ is defined as follows. If $(i, j) \in E$, then $l_{ij} = -1$; if $(i, j) \notin E$ and $i \neq j$, then $l_{ij} = 0$; if $i = j$, then $l_{ii} = -\sum_{j \neq 1} l_{ij}$. The goal of pinning control is to drive the system from any initial state to the target state in finite time by pinning some selected nodes. Analogous to the node percolation, all nodes are ranked in the descending order by a given index. Then, we successively pin nodes one by one according to the ranking and quantify the synchronizability of the pinned networks, which can be measured by the reciprocal of the smallest nonzero eigenvalue of the principal submatrix (Liu et al., 2018; Pirani and Sundaram, 2016) (a smaller value corresponds to a higher synchronizability), namely $1/\mu_1(L_{-Q})$, where $Q$ is the number of pinned nodes, $L_{-Q}$ is the principal submatrix, obtained by deleting the $Q$ rows and columns corresponding to the $Q$ pinned nodes from the original Laplacian matrix $L$, and $\mu_1(L_{-Q})$ is the smallest nonzero eigenvalue of $L_{-Q}$. Inspired by the metric Robustness, we propose a similar metric named pinning efficiency to characterize the performance of an index subject to pinning control, as

$$P = \frac{1}{Q_{max}} \sum_{Q=1}^{Q_{max}} \frac{1}{\mu_1(L_{-Q})}, \tag{4}$$

where $Q_{max}$ is the maximum number of pinned nodes under simulation. Here we set $Q_{max} = 0.3N$, and we have checked that the choices of $Q_{max}$ will not affect the conclusion. Fig. 5 shows how $1/\mu_1(L_{-Q})$ decays with increasing number of pinned nodes. Obviously, a faster decay corresponds to a better performance. Table 3 compares the pinning efficiency of the six indices. Overall speaking, betweenness and articulation ranking are better than cycle ratio, and cycle ratio is better than degree, H-index and coreness.

**Table 2. The robustness *R* of the six indices on the six real networks.**

| Network | Degree | H-index | Coreness | Cycle ratio | Betweenness | Articulation ranking |
|---|---|---|---|---|---|---|
| **C. elegans** | 0.3303 | 0.3678 | 0.3778 | **0.3167** | 0.3481 | 0.4443 |
| **Email** | **0.2511** | 0.2813 | 0.2949 | 0.2597 | 0.2578 | 0.3316 |
| **Jazz** | 0.4394 | 0.4479 | 0.4546 | 0.4190 | **0.3955** | 0.4514 |
| **NS** | 0.0539 | 0.1173 | 0.1803 | 0.0536 | **0.0488** | 0.0956 |
| **USAir** | 0.1236 | 0.1487 | 0.1587 | 0.1312 | **0.1129** | 0.3203 |
| **Yeast** | 0.1960 | 0.2630 | 0.2901 | 0.1726 | **0.1437** | 0.2491 |

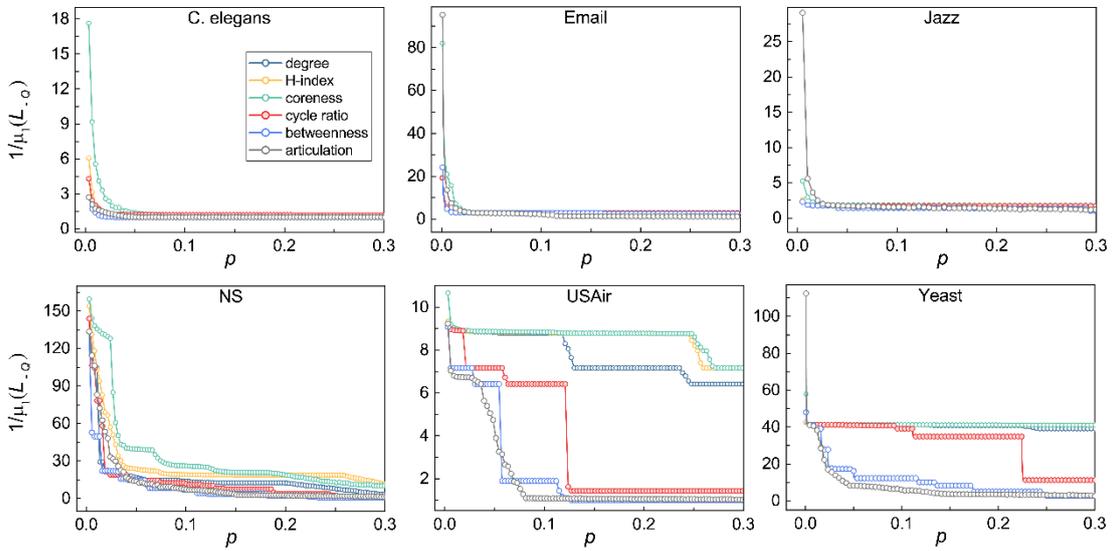

**Fig. 5. The performance of the six indices on pinning control.** The *y*-axis shows the synchronizability after pinning a fraction of nodes, and the *x*-axis denotes the ratio of pinned nodes.

**Table 3. The pinning efficiency *P* of the six indices on the six real networks.**

| Network | Degree | H-index | Coreness | Cycle ratio | Betweenness | Articulation ranking |
|---|---|---|---|---|---|---|
| **C. elegans** | 1.2614 | 1.2938 | 1.6490 | 1.2637 | **1.0368** | 1.0724 |
| **Email** | 3.1273 | 3.1445 | 4.0391 | 3.1377 | 2.9635 | **2.5954** |
| **Jazz** | 1.7928 | 1.8324 | 1.9368 | 1.7533 | **1.4341** | 2.0497 |
| **NS** | 16.125 | 25.8633 | 32.1256 | 12.9024 | **8.4350** | 11.7224 |
| **USAir** | 7.6831 | 8.5382 | 8.6007 | 3.6804 | 2.2392 | **2.0640** |
| **Yeast** | 40.670 | 41.2341 | 41.2826 | 31.1160 | 10.4713 | **7.4518** |

Lastly, we consider the spreading dynamics. Since in viral marketing and online information transmission, people are more interested in maximizing the reach in short time, and in epidemiological control, the most critical issue is the spreading range and control measures in the early stage of outbreak (e.g., see the discussion of the efficacy of early control measures for COVID-19 (Liu et al., 2020; Chen and Zhou, 2021)), we concentrate on the fast influencers that play the dominant role in the early stage (Zhou et al., 2019). To quantify the influence of a set of selected nodes, we simulate the standard susceptible-infected-recovered (SIR) spreading dynamics (Pastor-Satorras et al., 2015), where at each time step, each susceptible node will be infected by an infected neighbor with probability $\beta$, and each infected node will be recovered with probability $\gamma$. Initially, the top-$0.1N$ nodes selected by each index are set to be infected and others are susceptible. The indices are ranked by cumulative infected nodes at a certain time step $t$, the more the better. Here we consider the case at $\beta = \beta_c$ and $\gamma = 1$, where

$$\beta_c = \langle k \rangle / (\langle k^2 \rangle - \langle k \rangle) \tag{5}$$

is the spreading threshold (Castellano and Pastor-Satorras, 2010; Pastor-Satorras et al., 2015) when $\gamma = 1$. Here $\langle k \rangle$ and $\langle k^2 \rangle$ are the average degree and the average squared degree, respectively. Fig. 6 reports the rankings of the six indices at time steps $t = 1$, $t = 2$, $t = 4$ and $t = 8$, which are averaged over 1000 independent runs. The best-performed index is ranked No. 1, the runner up is ranked No. 2, …, and the worst one is ranked No. 6. Among the 24 matches (i.e., 6 networks and 4 time steps), cycle ratio gets ranked No. 1 for 14 times, and No. 2 for 8 times, with only 2 times being ranked No. 3. Overall speaking, it outperforms other indices. The results for more $(\beta, t)$ parameter sets are presented in Supplementary Information Section IV.

In addition to real networks, we have also analyzed two types of synthetic networks, the Erdős-Rényi (ER) networks (Erdős and Rényi, 1960) and Barabási-Albert (BA) networks (Barabási and Albert, 1999). The overall performance of cycle ratio is just in the middle of the six indices. The not-so-good performance may be resulted from the lack of short cycles in both ER and BA networks (partially reflected by the small clustering coefficient), since usually the longer cycles correspond to weaker relationships among associated nodes (Katz, 1953; Lü and Zhou, 2011). Detailed results about ER and BA networks are presented in the Supplementary Information Section V.

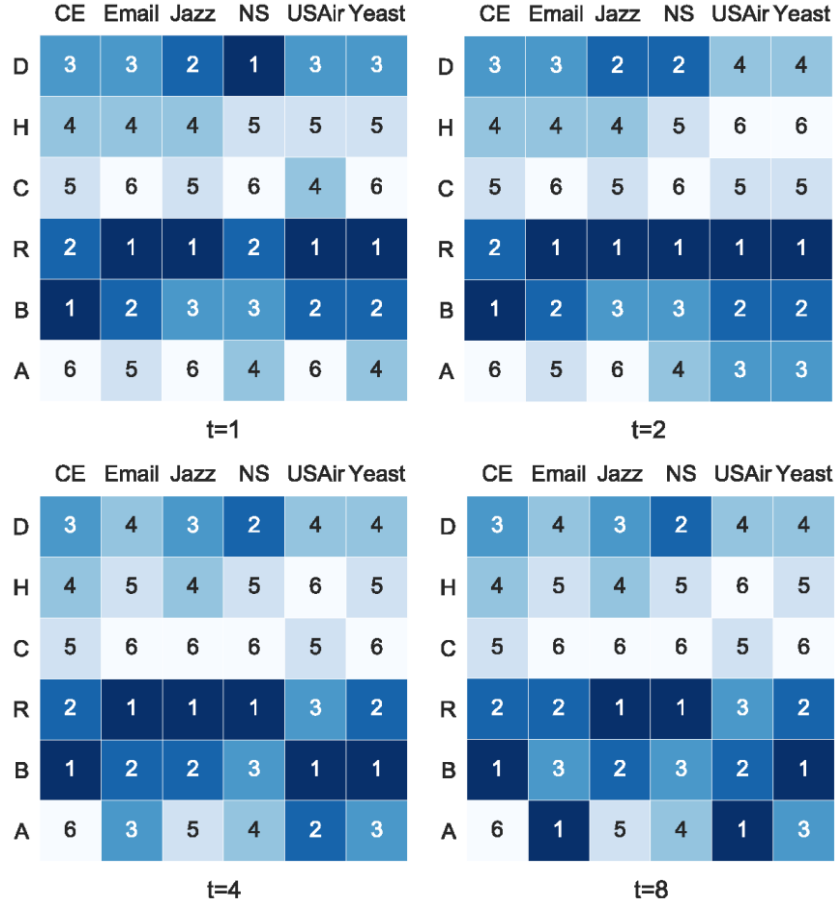

**Fig. 6. The performance of the six indices on spreading dynamics.** Each matrix presents the results of comparison of the six indices in a time step, where D, H, C, R, B and A represent degree, H-index, coreness, cycle ratio, betweenness and articulation ranking, respectively. CE is the abbreviation of C. elegans. The elements in each matrix are the rankings of six indices at the corresponding time step, which are all visualized by the color: the better the deeper. The infection probability is set as $\beta = \beta_c$ for each network.

**Discussion**

To represent cycle information of a network, this paper defines a novel matrix, called cycle number matrix, with which a new index, called cycle ratio, can be calculated to quantify the importance of an individual node by simply measuring to which extent it is involved in other nodes' associated shortest cycles. The basic idea underlying such an index is that if cycles are important in maintaining connectivity and interacting dynamics, then a node involved in many cycles should be vital. Experiments on real networks show that cycle ratio and betweenness outperform the other indices in identifying vital nodes that are critical in maintaining the network connectivity, efficient in pinning control and influential in epidemic spreading. Our finding thus has potential applicability in practice. For the node percolation, the top-ranked nodes should be firstly protected to maintain the network connectivity if there is a risk of functional loss of nodes. Reversely, if one would like to initiate an intentional attack, the top-ranked nodes are considered to be the primary targets. Such scenario is relevant to power grids (Albert et al., 2004), air transportation networks, financial networks (Haldane and May, 2011), Internet, and so on (Li et al., 2021). Note that, when we consider an attack to an airport in the modern society, it does not mean we need to physically destroy it but disturb its information systems and signal systems. The critical nodes in pinning control can be pinned to efficiently approach the consensus of multiple agents (Chen et al., 2009) and to ensure the coordination of unmanned aerial vehicles (Tang et al., 2012) and mobile sensor networks (Ögren et al., 2004). Lastly, cycle ratio is a very good index to find the susceptible individuals that need to be vaccinated in the early stage of epidemic spreading (Pei and Makse, 2013; Zhou et al., 2019).

It's worth noting that the performance of cycle ratio is not necessarily better if longer cycles are considered. This is because when the longer cycles are counted, the difference in local cycle structure might be depressed. That is to say, the sets of associated cycles of many nodes will become more similar (i.e., with larger overlap), which may eventually lead to the decrease of the discriminability and thus the accuracy of the cycle ratio (see Supplementary Information Section VI).

An obvious insufficiency of cycle ratio is that it could not be applied for trees or tree-like networks. Even for normal networks, a fraction of nodes may not be associated with any cycles. These nodes' influences may be different but they are all assigned the same cycle ratio zero. One straightforward way to solve this issue is to combine cycle ratio with some other indices, for example, a mixed index could be $r^* = r_i + \varepsilon k_i$ with $\varepsilon$ being a tunable parameter, hence all nodes with zero cycle ratio can be ranked by their degrees. Since cycle ratio and degree will produce remarkably different rankings, a subtly designed combination of cycle ratio and degree has the potential to generate much better results than the single index. Similar improvement could also be achieved by combining cycle ratio with H-index or coreness. In contrast, the expected improvement by combining degree, H-index and coreness is lower since they are already very similar to each other. We leave this detailed problem for future study.

In addition, the method used to characterize the cycle structure can be extended to deal with hypernetworks (Suo et al., 2018), where a hyperedge represents the interaction between multiple nodes. Treating hyperedges as the cycles in the set $S$ and denoting $\Omega$ the incidence matrix, whose element $\Omega_{ie}$ indicates whether node $i$ belongs to hyperedge $e$ ($\Omega_{ie} = 1$ indicates the belongness and $\Omega_{ie} = 0$ otherwise), then we can obtain a matrix similar to the cycle number matrix by multiplying the incidence matrix by its transposed matrix, say $\Omega\Omega^T$. Therefore, we can quantify a node's importance in a hypernetwork by its participation to other nodes' hyperedges.

We end this paper with two open issues. Firstly, analogous to cycle ratio, one may also design cycle-based indices to quantify the likelihood of the existence of any unobserved link, which can find applications in solving the link prediction problem. Secondly, the good performance of cycle ratio, as well as the lower correlations between cycle ratio and other benchmark centralities, encourages the in-depth studies on cycle structure. As shown in Supplementary Information Section VII, none of degree-preserved null model (Maslov and Sneppen, 2002), Watts-Strogatz model (Watts and Strogatz, 1998) and Barabasi-Albert model (Barabási and Albert, 1999) can well reproduce the cycle-based statistics of real networks, indicating that the understanding about how cycles are formed may unfold novel mechanisms underlying network organization. In addition to the shortest cycles, higher-order cycles also play important roles in network structure and functions (Sizemore et al., 2018; Shi et al., 2019). Thus We expect to find more insights from analyzing longer and higher-order cycles in the future with the help of methodologies from algebraic topology (Mahadevan et al., 2006; Shi et al., 2019) and sufficient computational resource, and extend the findings and scope of applications reported in this paper.

**Resource Availability**

*Lead Contact* Further information and requests for resources and reagents should be directed to and will be fulfilled by the Lead Contact, Linyuan Lü (linyuan.lv@gmail.com).
*Data and Code Availability* Data used in this work are provided as Supporting Data and can also be accessed at http://linkprediction.org/index.php/link/resource/data/1.

**Methods**

**Degree, H-index and Coreness**. Degree of a node is the number of its immediate neighbors. H-index of a node $i$ is the maximum integer $h$ such that there are at least $h$ neighbors of node $i$ with degrees no less than $h$. Coreness is obtained by the k-core decomposition (Kitsak et al., 2010). The k-core decomposition process starts by removing all nodes with degree $k = 1$. This may cause new nodes with degree $k \leq 1$ to appear. These are also removed and the process stops when all remaining nodes are of degree $k > 1$. The removed nodes and their associated links form the 1-shell, and the nodes in the 1-shell are assigned a coreness value 1. This pruning process is repeated to extract the 2-shell, that is, in each step the nodes with degree $k \leq 2$ are removed. Nodes in the 2-shell are assigned a coreness value 2. The process is continued until all higher-layer shells have been identified and all nodes have been removed. In the literature, coreness is also referred to as k-shell index.

**Betweenness**. Betweenness is defined as the fraction of shortest paths between node pairs that pass through the node of interest (Brandes, 2001; Zhou et al., 2006). The betweenness centrality of an arbitrary node $v$ is

$$C_B(v) = \sum_{s \in V, t \in V, s \neq v \neq t} \frac{\sigma_{st}(v)}{\sigma_{st}}, \tag{6}$$

where $\sigma_{st}$ is the number of shortest paths between nodes $s$ and $t$, and $\sigma_{st}(v)$ is the number of shortest paths between $s$ and $t$ which pass through node $v$.

**Articulation ranking**. Articulation ranking of a node is defined as the number of components that would be added to the network upon the deletion of that node (Tishby et al., 2018).

**Kendall's Tau**. We consider any two indices associated with all $N$ nodes, $X =$

$(x_1, x_2, \ldots, x_N)$ and $Y = (y_1, y_2, \ldots, y_N)$, as well as the $N$ two-tuples $(x_1, y_1), (x_2, y_2), \ldots, (x_N, y_N)$. Any pair $(x_i, y_i)$ and $(x_j, y_j)$ are concordant if the ranks for both elements agree, namely if both $x_i > x_j$ and $y_i > y_j$ or if both $x_i < x_j$ and $y_i < y_j$. They are discordant if $x_i > x_j$ and $y_i < y_j$ or if $x_i < x_j$ and $y_i > y_j$. Here $n_+$ and $n_-$ are used to represent the number of concordant and discordant pairs, respectively. In addition, $t_X$ is the number of the pairs in which $x_i = x_j$ and $y_i \neq y_j$, and $t_Y$ is the number of the pairs in which $x_i \neq x_j$ and $y_i = y_j$. Notice that If $x_i = x_j$ and $y_i = y_j$, the pair is not added to either $t_X$ or $t_Y$. Comparing all $N(N-1)/2$ pairs of two-tuples, the Kendall's Tau is defined as (Knight, 1966)

$$\tau = \frac{(n_+ - n_-)}{\sqrt{(n_+ + n_- + t_X)} \times \sqrt{(n_+ + n_- + t_Y)}}. \qquad (7)$$

If $X$ and $Y$ are independent, $\tau$ should be close to zero, and thus the extent to which $\tau$ exceeds zero indicates the strength of correlation. The above definition of Kendall's Tau (Knight, 1966) is an improved version of the original definition (Kendall, 1938), specifically designed to deal with the case with many equivalent elements.

**Acknowledgments.** This work is supported by the National Natural Science Foundation of China (Nos. 11622538, 61673150, 61433014, 11975071), and the Zhejiang Provincial Natural Science Foundation of China (No. LR16A050001). L.L. and T.Z. acknowledges the Science Strength Promotion Programme of UESTC.

**Author contributions.** T.F., L.L. and D.S. conceived the idea and designed the experiments and T.Z. provided a complement to the design. T.F. and L.L. performed the research. All authors analyzed the data. T.F., L.L. and T.Z. wrote the paper. D.S. edited this paper. All authors discussed the results and reviewed the manuscript.

# Supplementary Information for
Characterizing cycle structure in complex networks

## I. The number of cycles with different lengths

Fig. S1 shows the number of cycles with different lengths. One can observe an exponential growth.

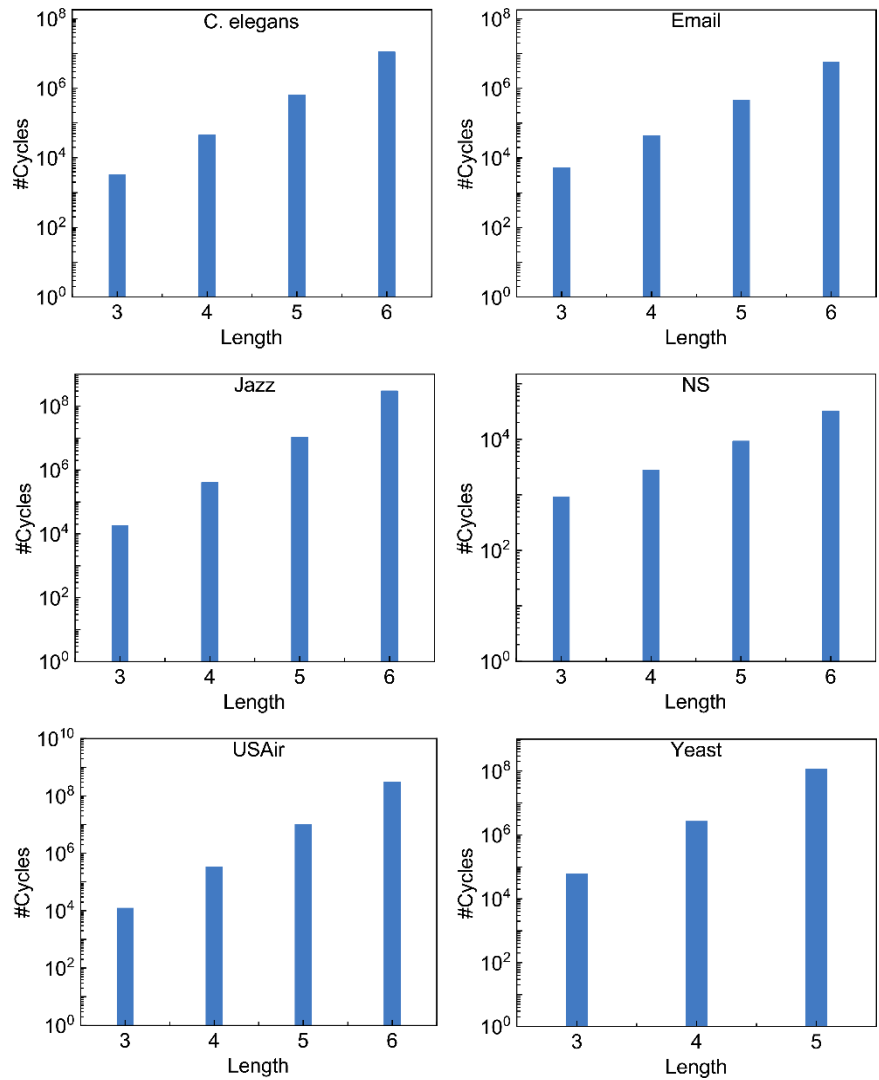

**Fig. S1. The number of cycles with different lengths in log-linear plot.**

## II. Correlations between index pairs

Fig. S2 shows the correlation matrices of the six indices for the six real networks under consideration. One can observe that the correlations among degree, H-index and coreness are very high, while the correlations between cycle ratio and others are much lower.

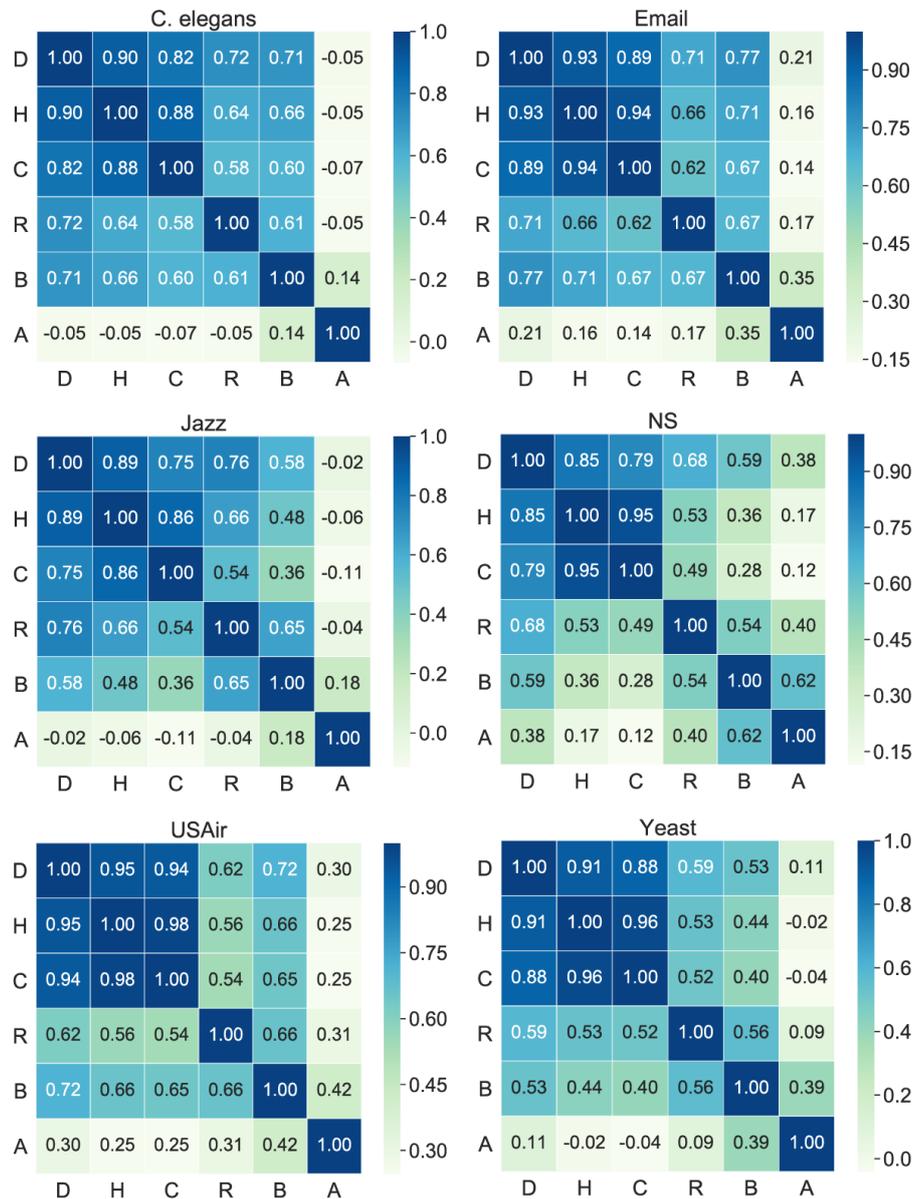

**Fig. S2. The correlation matrices of the six indices for the six networks.** Here D, H, C, R, B and A represent degree, H-index, coreness, cycle ratio, betweenness and articulation ranking, respectively. Each value is a Kendall's Tau between two indices for a real network. The value is visualized by the color.

## III. Distributions of indices

Figs. S3-S8 shows the distributions of the six indices on the six real networks, from which one can observe that the distinguishability of cycle ratio is good.

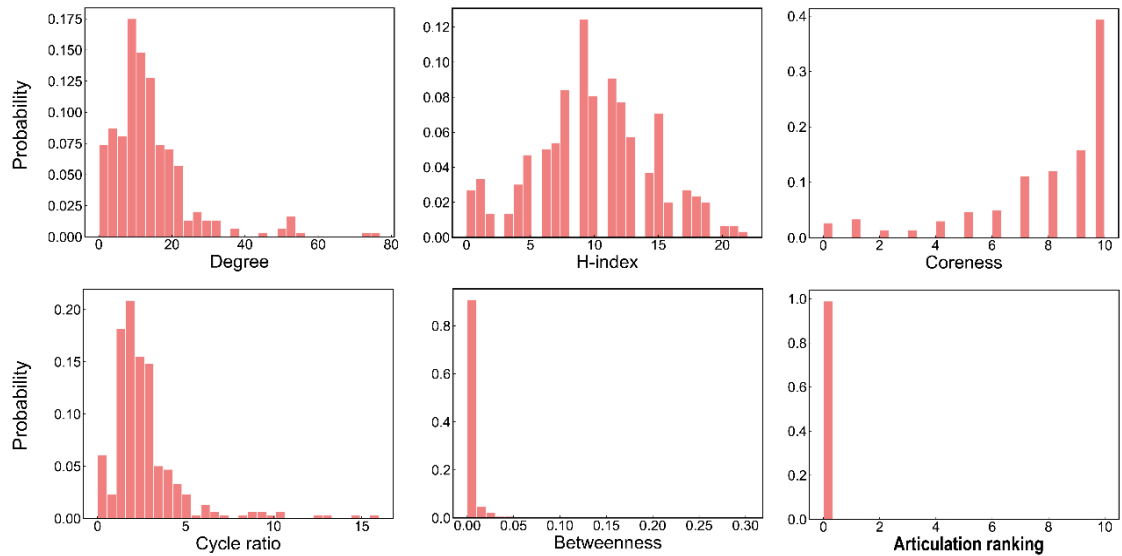

**Fig. S3. The histograms of the values of the six indices for C. elegans.**

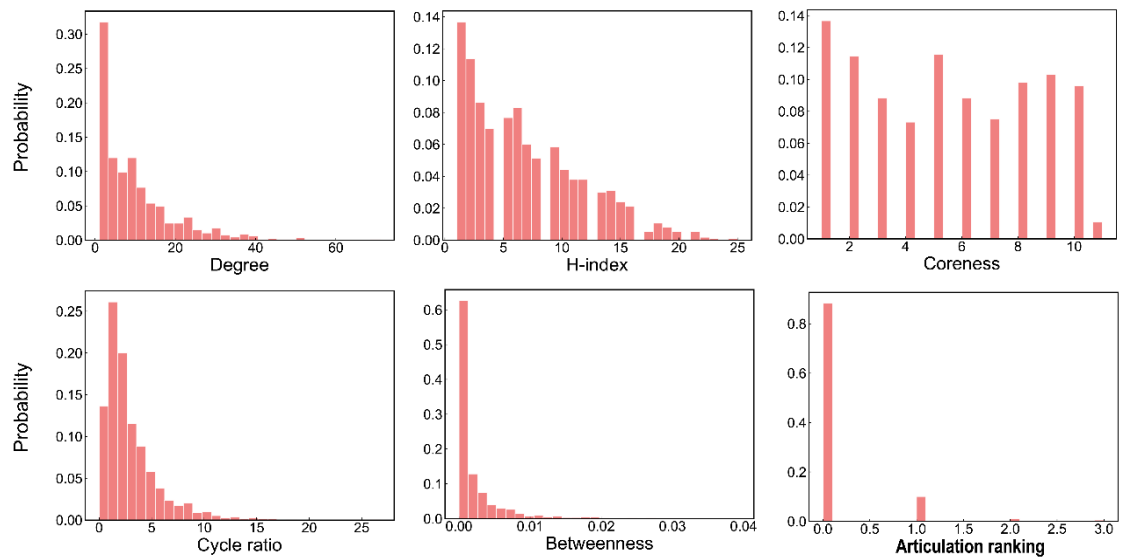

**Fig. S4. The histograms of the values of the six indices for Email.**

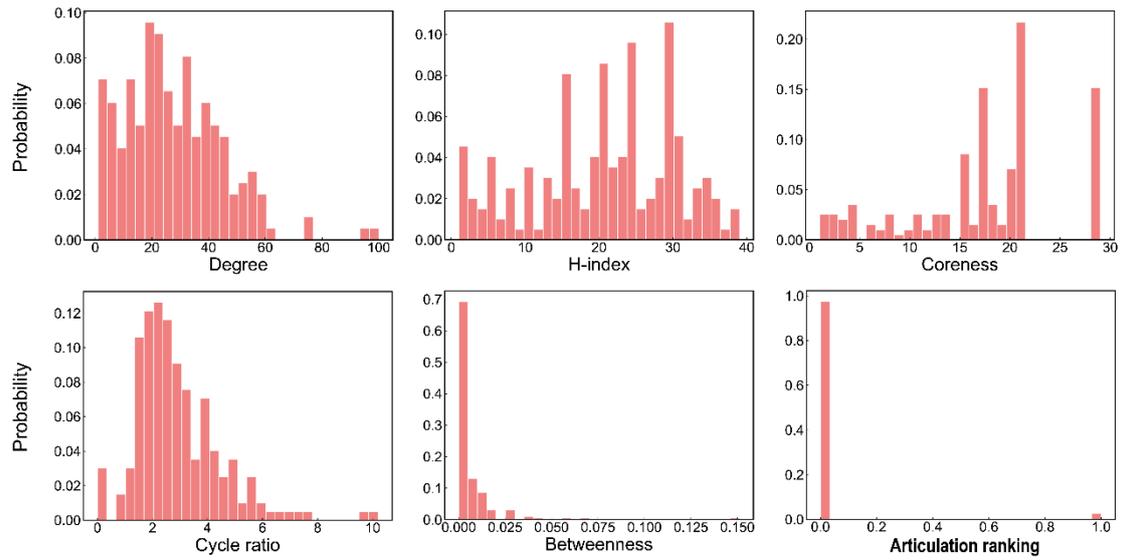

**Fig. S5.** The histograms of the values of the six indices for Jazz.

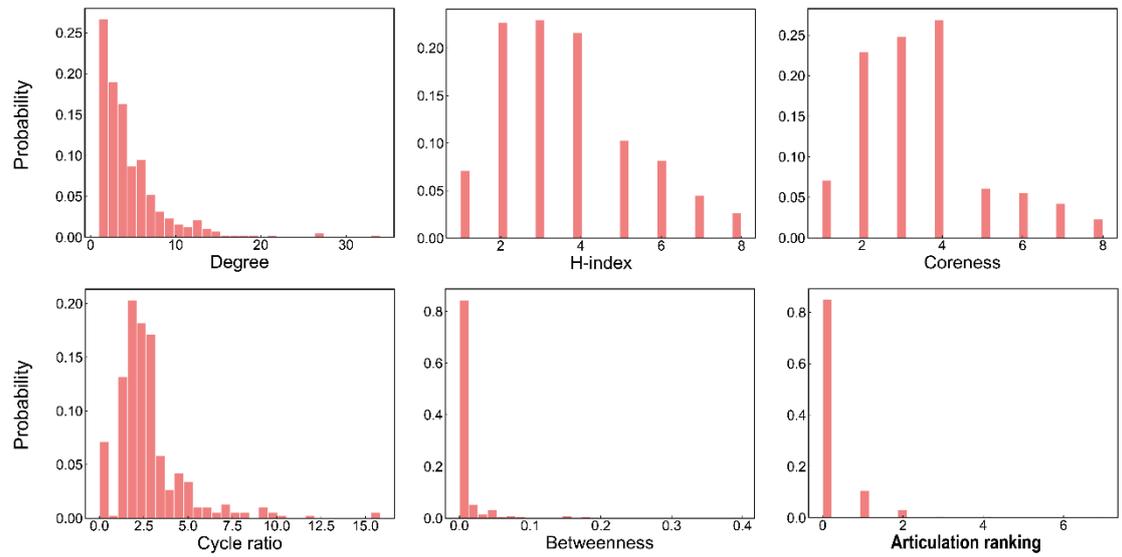

**Fig. S6.** The histograms of the values of the six indices for NS.

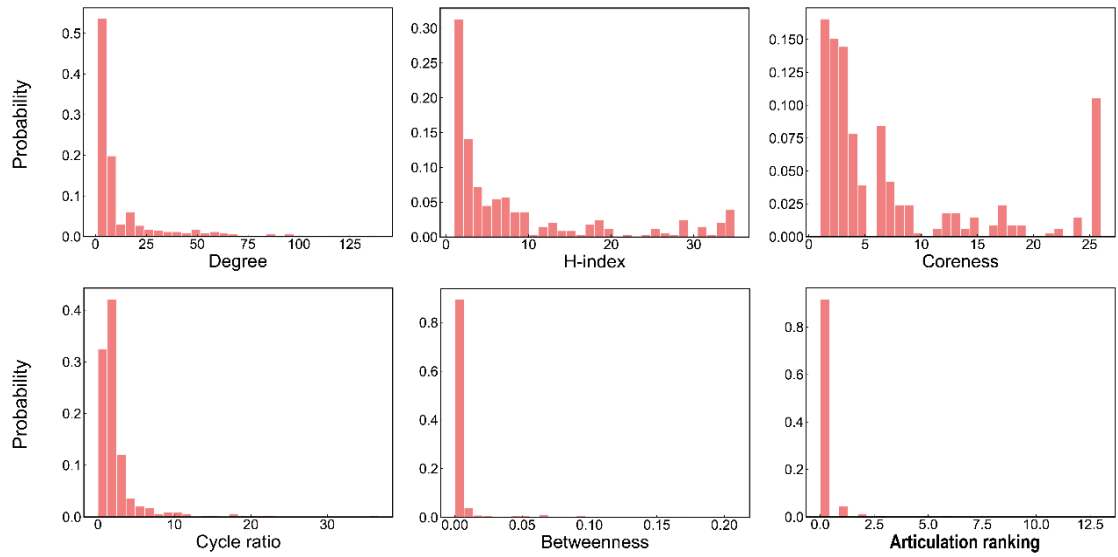

**Fig. S7. The histograms of the values of the six indices for USAir.**

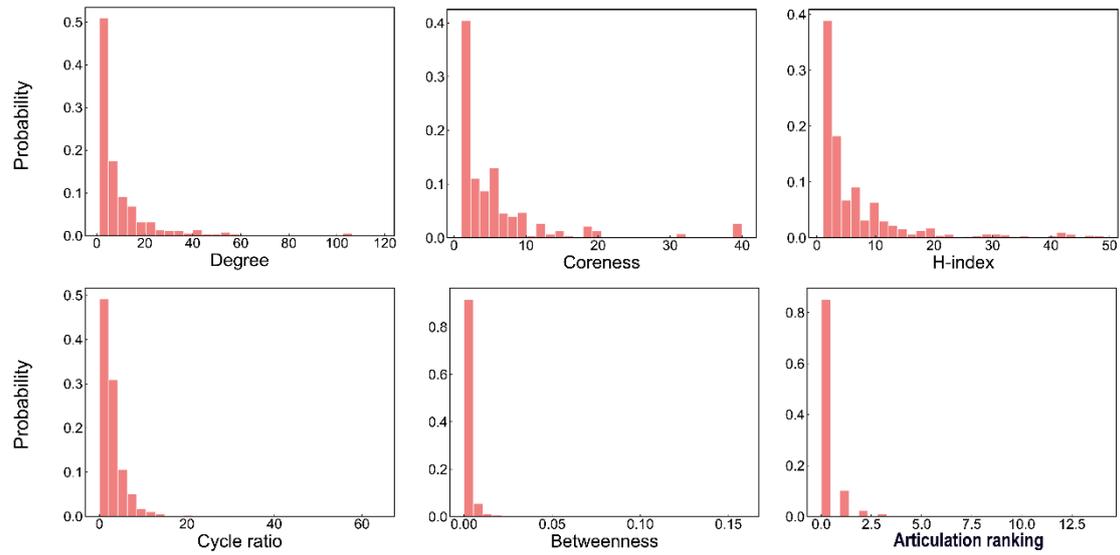

**Fig. S8. The histograms of the values of the six indices for Yeast.**

## IV. Detailed results for epidemic spreading

Fig. S9 shows the cumulative number of infected nodes with three infection probabilities $0.5\beta_c$, $\beta_c$ and $2\beta_c$ at the first 10 time steps, which are averaged over 1000 independent runs. Fig. S10 shows the rankings of the result in Fig. S9. Analogously, Fig. S11 and Fig. S12 show the results of the top-$0.05N$ nodes respectively.

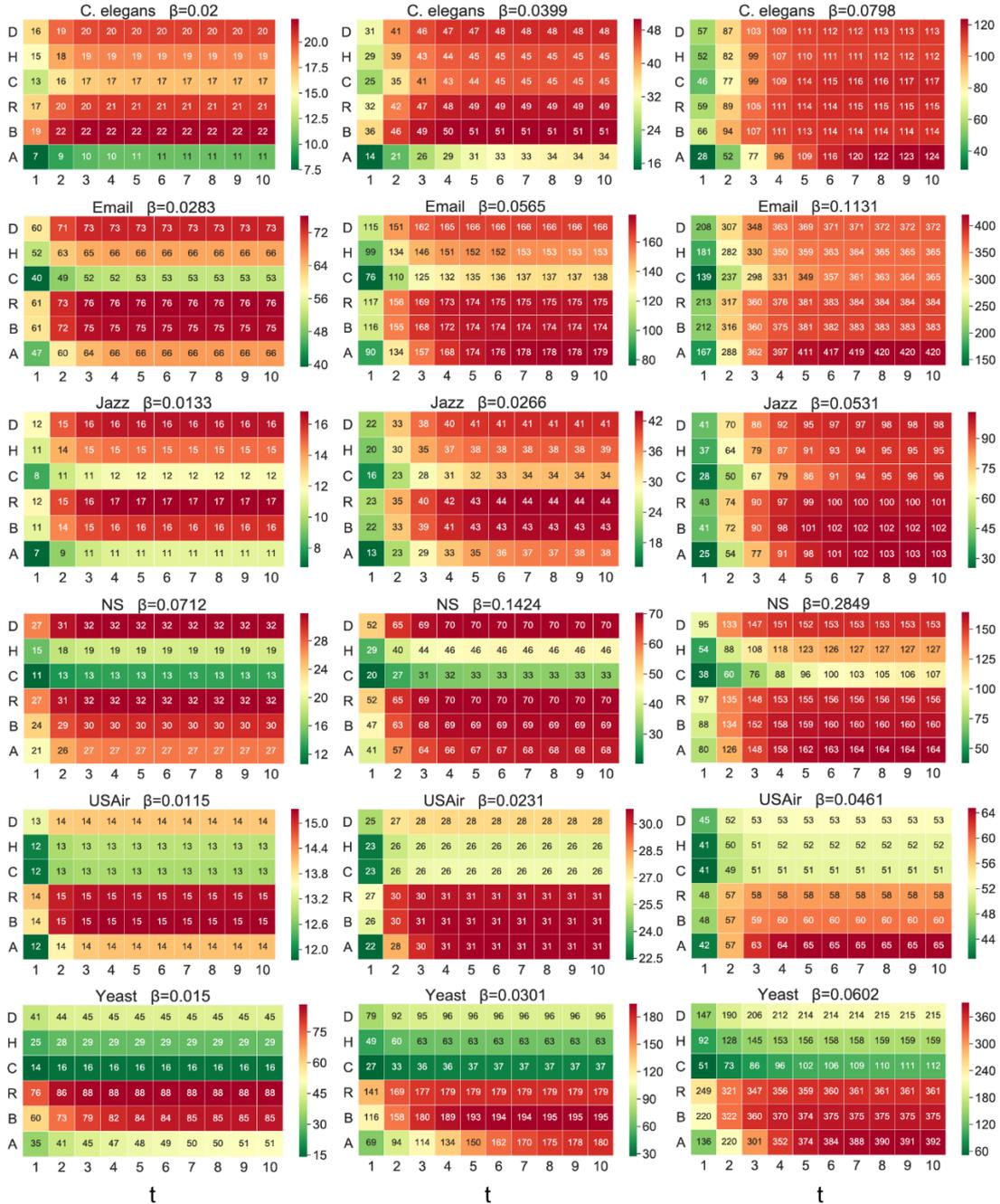

**Fig. S9．The cumulative number of infected nodes.** Each element is the cumulative number of infected nodes for the corresponding network with the top-$0.1N$ nodes being the initially infected seeds. The values are visualized by the color: the better the warmer.

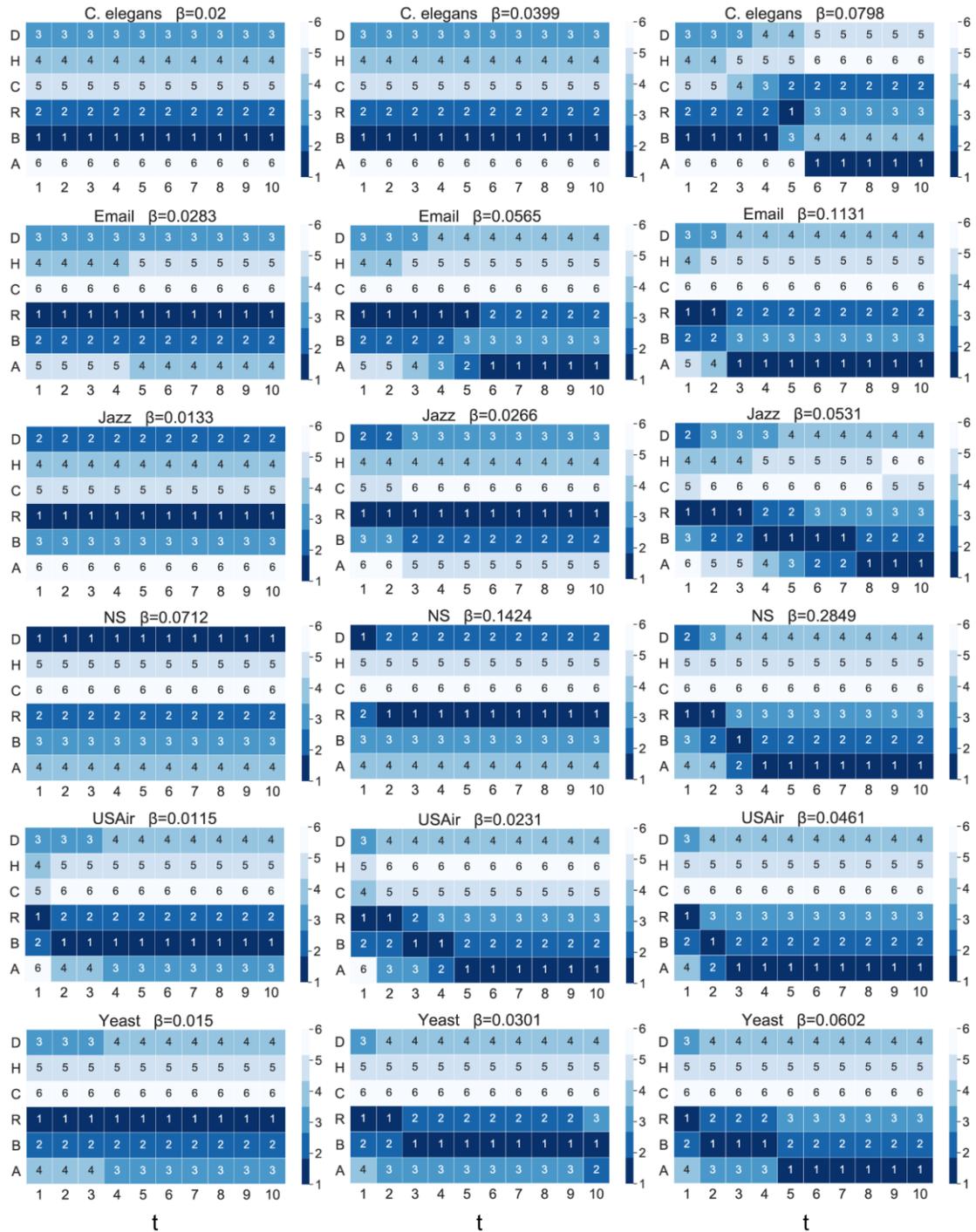

**Fig. S10．The performance of the six indices on spreading dynamics.** Each matrix presents the results of comparison of the six indices in the first 10 time steps. The parameter setting is the same as that for Fig. S9. The elements in each matrix are the rankings of six indices at the corresponding time steps, which are all visualized by the color: the better the deeper.

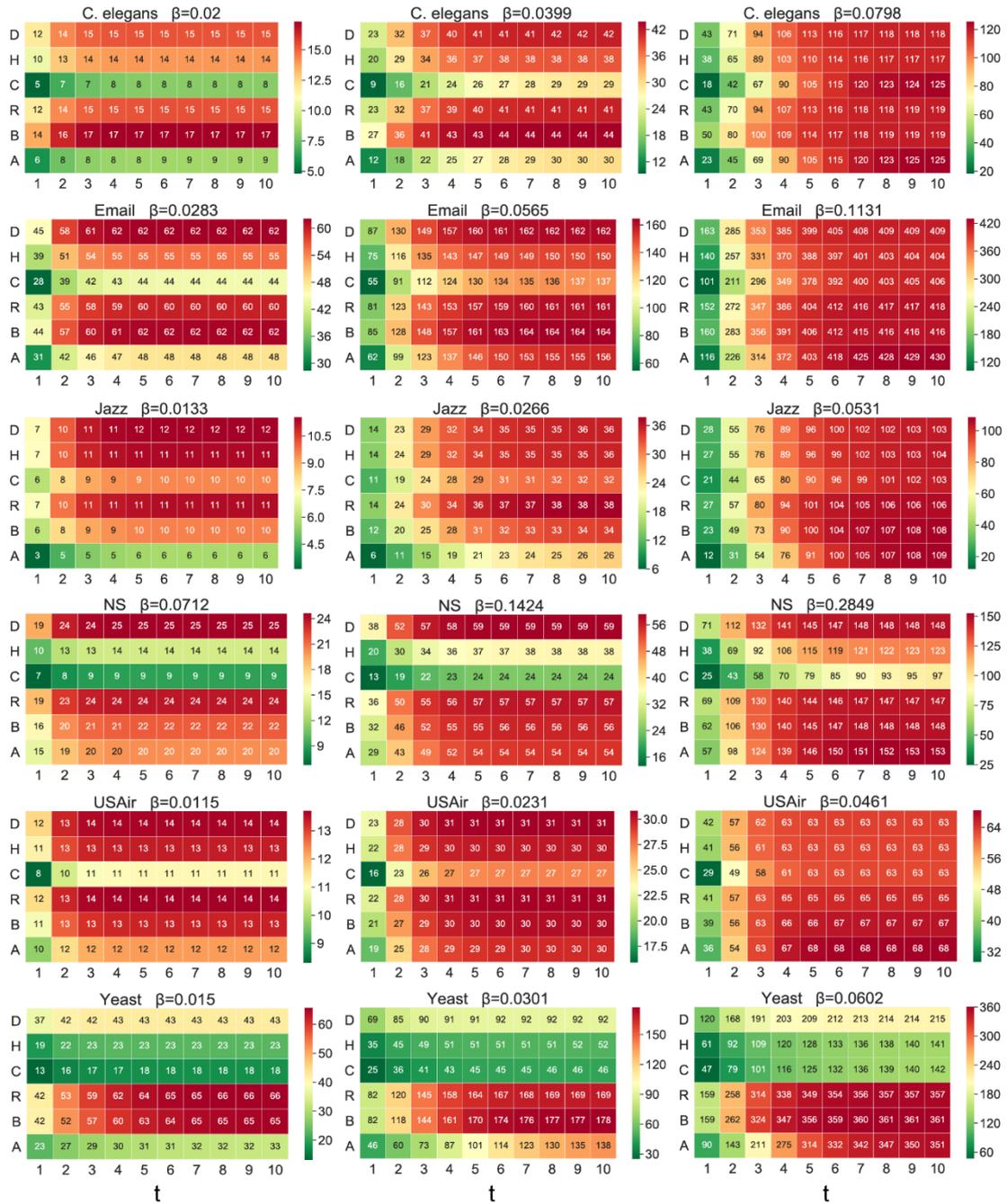

**Fig. S11. The cumulative number of infected nodes.** Each element is the cumulative number of infected nodes for the corresponding network with the top-0.05$N$ nodes being the initially infected seeds. The values are visualized by the color: the better the warmer.

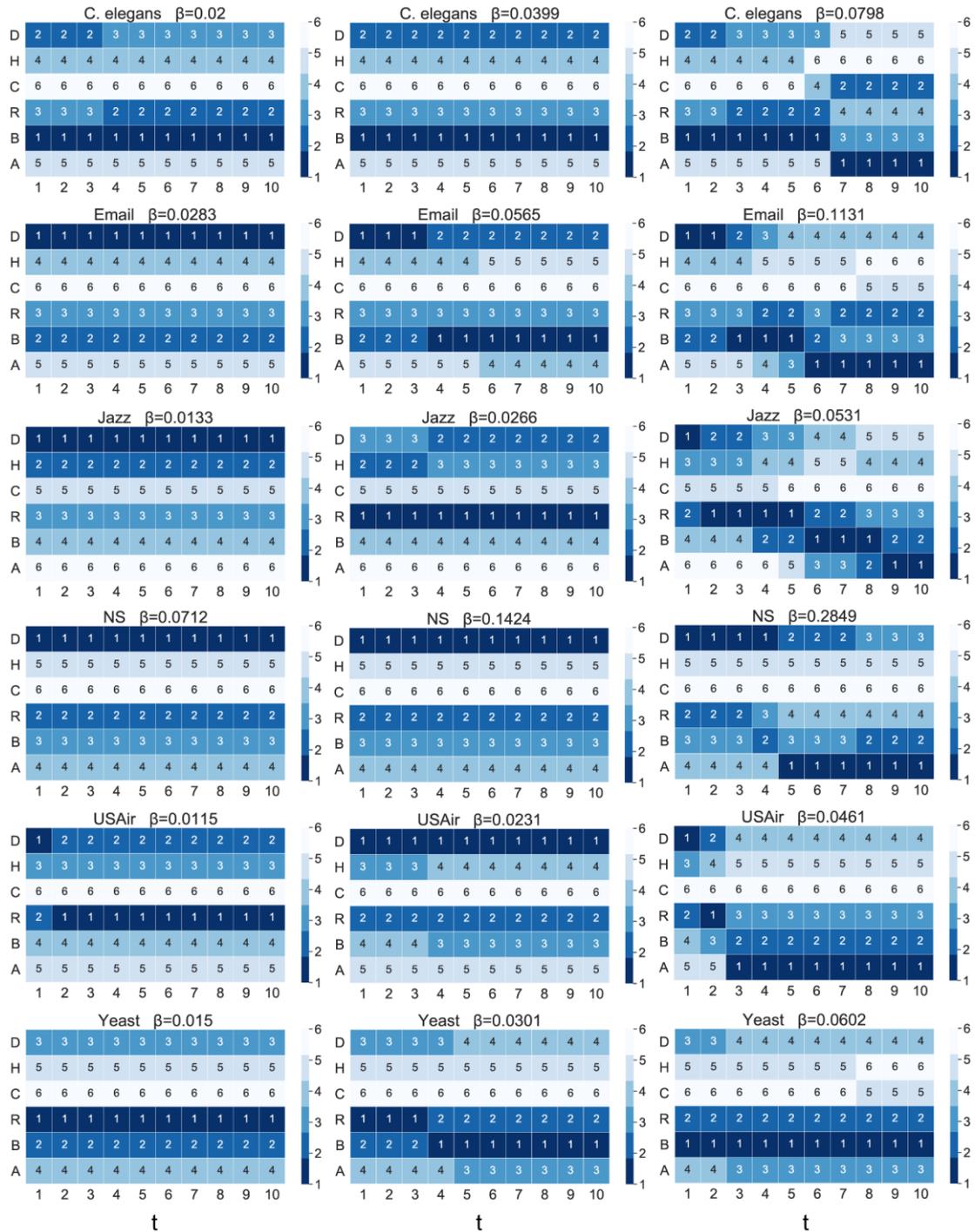

**Fig. S12. The performance of the six indices on spreading dynamics.** Each matrix presents the results of comparison of the six indices in the first 10 time steps. The parameter setting is the same as that for Fig. S11. The elements in each matrix are the rankings of six indices at the corresponding time steps, which are all visualized by the color: the better the deeper.

## V. Results for synthetic networks

Here we have analyzed two types of synthetic networks, the Erdos-Renyi (ER) networks and Barabasi-Albert (BA) networks. Table S1 presents the basic topological features of the two synthetic networks used for simulation. Primary results are shown in the following figures and tables: Fig. S13 and Table S2 for node percolation, Fig. S14 and Table S3 for pinning control and Fig. S15 for epidemic spreading.

**Table S1. The basic topological features of ER and BA networks.** Here $N$ and $M$ are the number of nodes and links, $\langle k \rangle$ and $\langle L \rangle$ are the mean degree and mean shortest distance, and $C$ is the clustering coefficient.

| Network | $N$ | $M$ | $\langle k \rangle$ | $\langle L \rangle$ | $C$ |
|---|---|---|---|---|---|
| **ER** | 1000 | 5000 | 10 | 3.25 | 0.01 |
| **BA** | 1000 | 4975 | 9.95 | 2.98 | 0.04 |

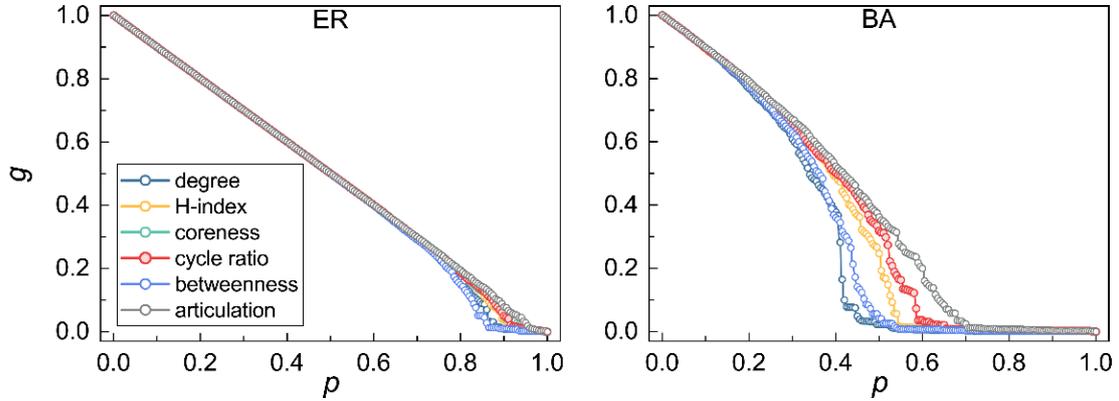

**Fig. S13. The performance of the six indices on node percolation for the two synthetic networks.** The $y$-axis shows the relative size of the largest component after node removal and the $x$-axis denotes the ratio of removed nodes.

**Table S2. The robustness $R$ of the six indices for the two synthetic networks.**

| Network | Degree | H-index | Coreness | Cycle ratio | Betweenness | Articulation ranking |
|---|---|---|---|---|---|---|
| **ER** | 0.4855 | 0.4919 | 0.4934 | 0.4927 | **0.4828** | 0.4963 |
| **BA** | **0.3074** | 0.3534 | 0.3959 | 0.3706 | 0.3211 | 0.3959 |

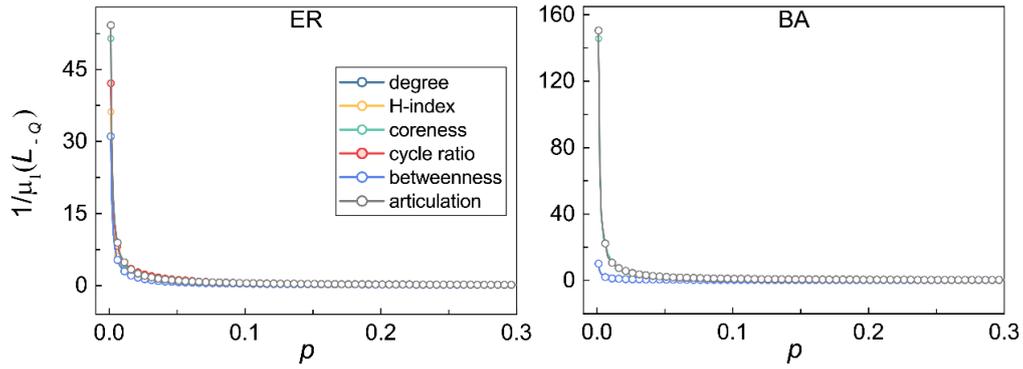

**Fig. S14. The performance of the six indices on pinning control for the two synthetic networks.** The *y*-axis shows the synchronizability after pinning a fraction of nodes, and the *x*-axis denotes the ratio of pinned nodes.

**Table S3. The pinning efficiency *P* of the six indices on the two synthetic networks.**

| Network | Degree | H-index | Coreness | Cycle ratio | Betweenness | Articulation ranking |
|---|---|---|---|---|---|---|
| **ER** | 0.7375 | 0.8325 | 1.0656 | 1.0917 | **0.7354** | 1.1159 |
| **BA** | 0.5534 | 0.5753 | 2.6814 | 0.5566 | **0.5524** | 2.6584 |

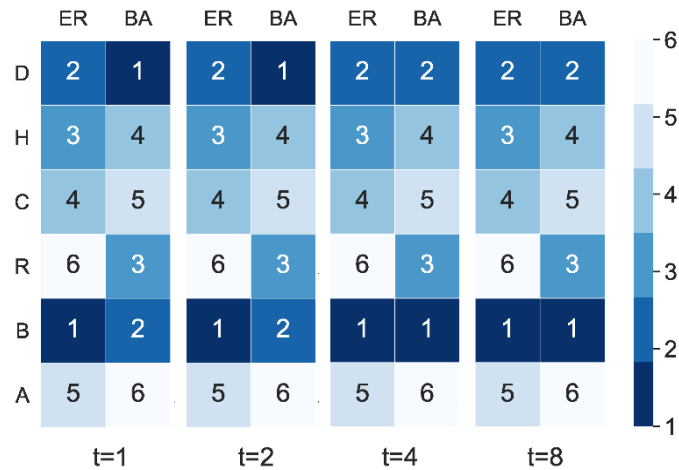

**Fig. S15. The performance of the six indices on spreading dynamics for ER and BA networks.** Each matrix presents the results of comparison of the six indices in a time step, where D, H, C, R, B and A represent degree, H-index, coreness, cycle ratio, betweenness and articulation ranking, respectively. Initially, the top-$0.1N$ nodes selected by each index are set to be infected. The elements in each matrix are the rankings of six indices at the corresponding time step, which are visualized by the color: the better the deeper. The infection probability is set as $\beta = \beta_c$ for each network.

## VI. Performance of longer cycles

Here we consider three different definitions of cycle ratios: (i) the original one involves only the shortest cycles in the set $S$; (ii) the second cycle ratio involves the set $S^{II}=\cup_{i \in V} S_i^{II}$, where $S_i^{II}$ denotes the set of cycles with length no more than the second shortest cycles of node $i$; (iii) the third cycle ratio involves the set $S^{III}=\cup_{i \in V} S_i^{III}$, where $S_i^{III}$ denotes the set of cycles with length no more than the third shortest cycles of node $i$. Since the calculation with longer cycles is highly time-consuming, we only test on two smaller networks. Fig. S16 and Table S4 show the results of node percolation, and Fig. S17 and Table S5 show the results of pinning control. One can see clearly from the results that the consideration of longer cycles will not necessarily improve the performance.

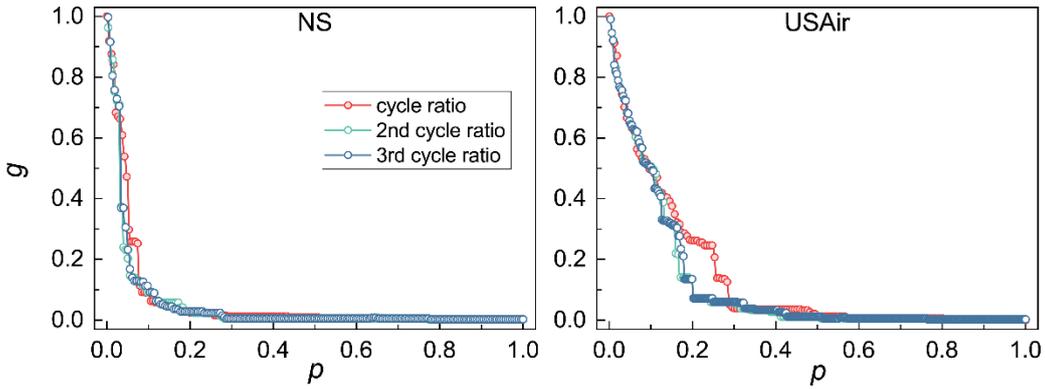

**Fig. S16. The performance of the three kinds of cycle ratios on node percolation for the two real networks.** The *y*-axis shows the relative size of the largest component after node removal and the *x*-axis denotes the ratio of removed nodes.

**Table S4. The robustness $R$ of the three kinds of cycle ratios for the two real networks.**

| Network | Cycle Ratio | Second Cycle Ratio | Third Cycle Ratio |
|---|---|---|---|
| **NS** | 0.0536 | **0.0471** | 0.0481 |
| **USAir** | 0.1312 | **0.1096** | 0.1115 |

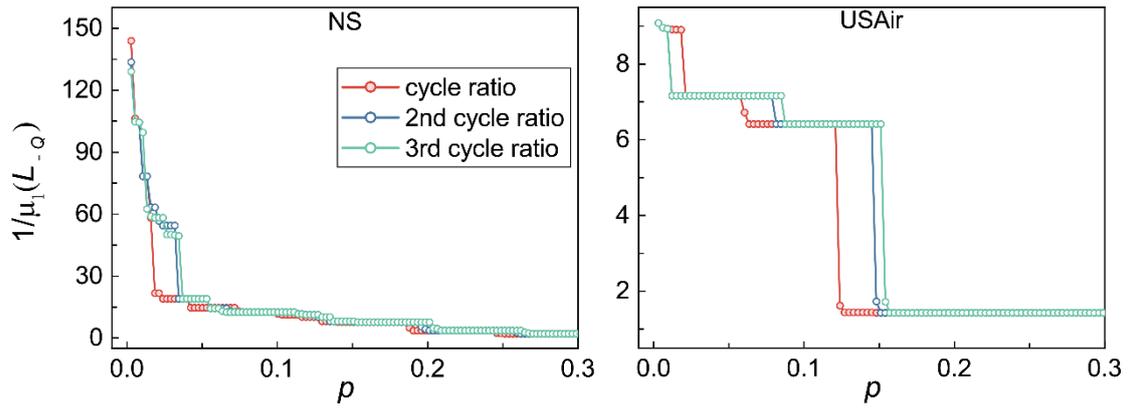

**Fig. S17. The performance of the three kinds of cycle ratios on pinning control for the two real networks.** The *y*-axis shows the synchronizability after pinning a fraction of nodes, and the *x*-axis denotes the ratio of pinned nodes.

**Table S5. The pinning efficiency *P* of the three kinds of cycle ratio for the two real networks.**

| Network | Cycle Ratio | Second Cycle Ratio | Third Cycle Ratio |
|---|---|---|---|
| **NS** | **12.9024** | 15.2354 | 15.4877 |
| **USAir** | **3.6804** | 4.1030 | 4.2186 |

## VII. Distributions of sizes of shortest cycles

We compare the distributions of sizes of shortest cycles in the set $S$ for real networks, corresponding null networks and two model networks with various parameters. Given a network, its degree-preserved null network can be obtained by repeatedly rewiring links. At each time step, two links ($a$, $b$) and ($c$, $d$) are randomly selected, if nodes $a$ and $d$ are not yet connected by a link and nodes $b$ and $c$ are not yet connected by a link, we remove links ($a$, $b$) and ($c$, $d$) from the current network and add two new links ($a$, $d$) and ($b$, $c$). Otherwise, we do nothing. Repeating such operation for sufficiently many times will lead to the null network that has exactly the same degree sequence with but is considered to be more random than the original network.

As shown in Fig. S18, the difference between the distributions of the sizes of shortest cycles in $S$ for real networks and their null networks is significant. For real networks, most cycles in $S$ are of size 3, while most cycles in $S$ for null networks are of size no less than 4.

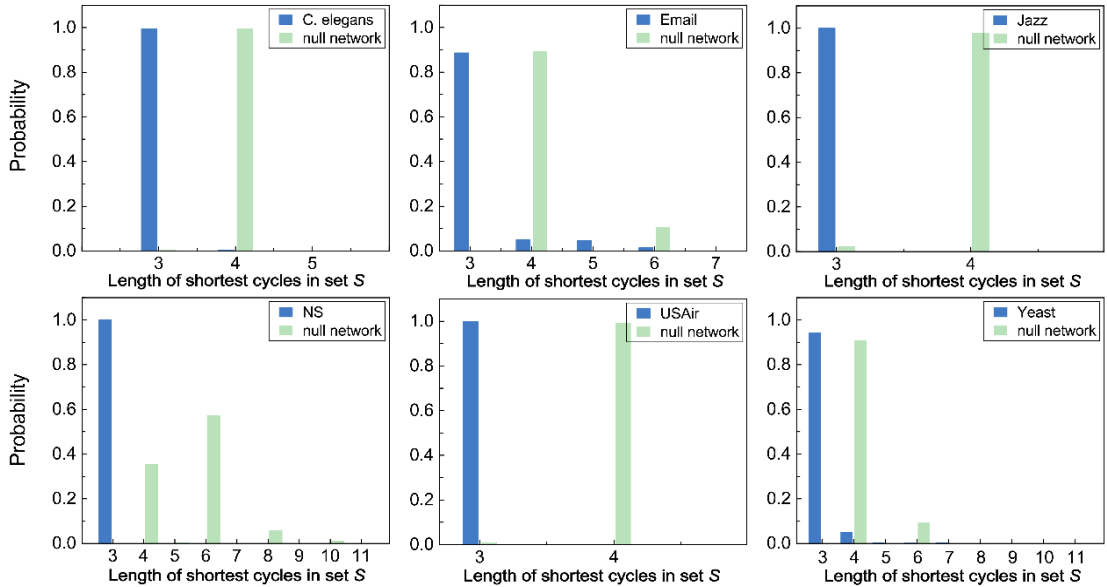

**Fig. S18. Comparison of distributions of shortest cycles' sizes in set $S$ for the six real networks and their corresponding degree-preserved null networks.**

Fig. S19 and Fig. S20 show the distribution of shortest cycles of the Watts-Strogatz (WS) networks and Barabasi-Albert (BA) networks, respectively. A WS network starts with a regular ring lattice where each node is connected to its $z$ nearest neighbors, and it is then obtained by rewiring each edge with probability $p$. A BA network starts with $m_0$ seed nodes. At each time step, a new node associated with $m$ edges will be added to the network and preferentially attached to the existed nodes with larger degree. It can be seen from Fig. S19 and Fig. S20 that the distributions of shortest cycles in these two classical models are significantly different from those in real networks.

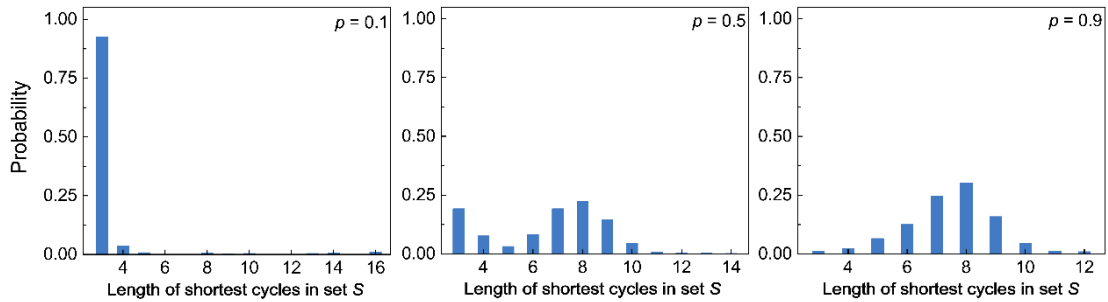

**Fig. S19. Distribution of lengths of shortest cycles in set $S$ of WS networks with different $p$.** The number of nodes and the connectivity are set as $N = 1000$, and $z = 4$.

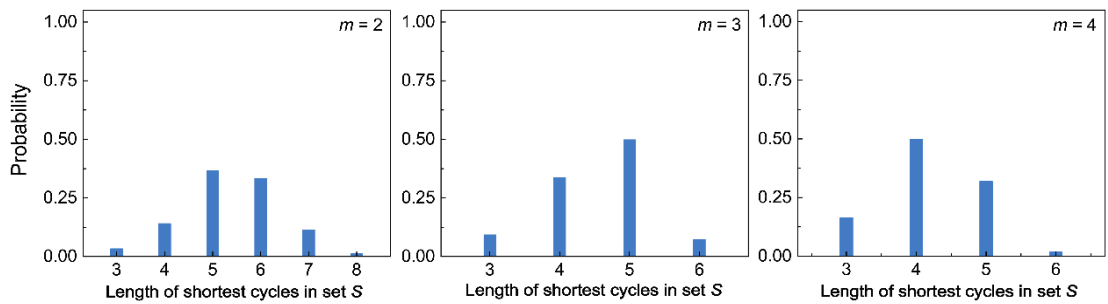

**Fig. S20. Distribution of lengths of shortest cycles in set $S$ of BA networks with different $m$.** The number of nodes and the seed size are set as $N = 1000$, and $m_0 = 5$.